\documentclass{taira}
\usepackage{tabularx}
\usepackage{graphicx}
\usepackage[square,sort,comma,numbers]{natbib}
\usepackage{pifont}
\usepackage{soul}
\usepackage{amsmath}
\usepackage{amssymb}
\usepackage{amsfonts}
\usepackage{color}
\usepackage{mathtools}
\usepackage{relsize}
\usepackage{wasysym}

\newcommand{\vol}{\mathop{\ooalign{\hfil$V$\hfil\cr\kern0.08em--\hfil\cr}}\nolimits}

\newcommand{\bs}[1]{\boldsymbol{#1}}

\newcommand{\abs}[1]{{\left\lvert#1\right\rvert}}
\newcommand{\norm}[1]{{\left\lVert#1\right\rVert}_2}

\newcommand{\mrm}[1]{\mathrm{#1}}

\title[Synchronization properties of the cylinder wake for periodic external forcings]{Phase-synchronization properties of laminar cylinder wake for periodic external forcings}

\author[M. A. Khodkar and K. Taira]{M. A. Khodkar$^1$ and Kunihiko Taira$^{1}$\thanks{Email addresses for correspondence: mkhodkar@ucla.edu}}

\affiliation{$^1$Department of Mechanical and Aerospace Engineering, University of California, Los Angeles, CA 90095, USA}

\date{\today}

\pubyear{2020}
\volume{??}
\pagerange{???--???}
\date{?; revised ?; accepted ?. - To be entered by editorial office}

\begin{document}

\maketitle

\begin{abstract}

We investigate the synchronization properties of the two-dimensional periodic flow over a circular cylinder using the principles of phase-reduction theory. The influence of harmonic external forcings on the wake dynamics, and the possible synchronization of the vortex shedding behind the cylinder to these forcings, is determined by evaluating the phase response of the system to weak impulse perturbations. These horizontal and vertical perturbations are added at different phase values over a period, in order to develop a linear one-dimensional model with respect to the limit cycle that describes the high-dimensional and nonlinear dynamics of the fluid flow via only a single scalar phase variable. This model is then utilized to acquire the theoretical conditions for the synchronization between the cylinder wake and the harmonic forcings added in the global near-wake region. Valuable insights are gained by comparing the findings of the present research against those rendered by the dynamic mode decomposition and adjoint analysis of the wake dynamics in earlier works. The present analysis reveals regions in the flow which enable phase synchronization or desynchronization to periodic excitations for applications such as active flow control and fluid-structure interactions.  

\end{abstract}



\section{Introduction \label{section:Intro}}

Dynamical systems with rythmic unsteadiness are omnipresent in engineering problems such as structural vibrations, electric circuits, chemical reactions and robotic locomotions \citep{Kuramoto1984, Pikovsky2001}. The nonlinear dynamics of such systems is characterized by a self-sustained oscillation, when the state of the system resides sufficiently close to its limit-cycle attractor. The rich physics of limit-cycle oscillators can abundantly be found in periodic fluid flows as well \citep{Huerre1990}. These unsteady flows have long been the subject of extensive research for the purpose of prediction, control and extracting the coherent fluid structures \citep{Holmes1996}. In particular, the study of phase-locking between the flow oscillation and harmonic actuations is essential to reveal the underlying synchronization mechanisms for modifying the dominant oscillation frequency of the flow in a mathematically rigorous and computationally inexpensive fashion \citep{Kawamura2013, Kawamura2015, Taira2018, Iima2019}. Within the scope of the present work, synchronization (phase-locking) is defined as the adjustment of the natural frequency of the flow to the frequency of the actuation. In the context of aerodynamic performance of flyers, altering the flow frequency can in turn lead to the enhancement of lift-to-drag ratio, e.g., by advancing or delaying the vortex shedding from the body \citep{Nair2020}. These objectives require the development of low-dimensional models describing the nonlinear interaction between all oscillators in the fluid system in a simplified fashion \citep{Nair2018}. 

Most control or reduction techniques conventionally employed for fluid flows rely on the linearization of the flow around a steady or quasisteady state, which makes them underperform when the deviation from the bifurcation point is relatively large \citep{Bewley2001}. Approaches such as Floquet theory \citep{Herbert1987} or phase-reduction theory \citep{Kuramoto2019} enable the extension of analysis to periodic flows, as they pave a path towards active flow control with respect to time-varying base flows. 

The present investigation provides a linear framework for the phase-based description of temporal evolution of time-periodic flows. In particular, we analyze the response of a two-dimensional (2D) circular cylinder wake to harmonic actuations. The Reynolds number of interest is adequately large so that the von K\`{a}rm\`{a}n vortex street forms, but it is also low enough to ensure that the wake remains laminar. This analysis proposes a novel and computationally tractable way for uncovering the regions in the flow best suited for the placement of external oscillatory forcing inputs, to achieve synchronization between the vortex shedding and these actuations, and consequently, to modify the shedding frequency. These regions are obtained by surveying the phase properties of the global near-wake region, significantly expanding the local study performed by \cite{Taira2018}. Furthermore, this methodology is capable of identifying the most sensitive regions of the flow to control inputs at any given time, enabling an energy-optimal control strategy, without any explicit knowledge of governing dynamics. Additional insights are gained by comparing the findings of the present phase-based analysis against those of other modal- and adjoint-based methods, especially by inspecting the spatial maps of synchronization at different harmonics. The paper is organized as follows. The phase-based, low-dimensional description of the flow is derived in \S \ref{section:Theory}. Section \ref{section:DNS} describes the numerical solver used to conduct the direct numerical simulations (DNS) of the flow. Section \ref{section:Results} presents the results of the phase-based model, specifically the regions optimal for synchronization and energy-efficient control, and validates the accuracy of the model predictions for synchronization against DNS results. This section also investigates the potential commonalities and the characteristic differences between the findings of the present model and the widely-used dynamic mode decomposition (DMD), while highlighting the desirable capabilities of the phase-based analysis. Section \ref{section:Conclusion} summarizes the main results, and discusses the outlook for subsequent studies.    

\section{Phase-based analysis of periodic flows \label{section:Theory}}

Consider a periodic flow whose dynamics is governed by
\begin{equation}
\dot{\bs{q}} = \bs{F}(\bs{q}) \, , \label{eqn:Dyn1}
\end{equation}
where $\bs{q}$ is the state vector of the flow with a stable limit cycle $\bs{q_0}$ such that $\bs{q_0}(\bs{x}, t+T) = \bs{q_0}(\bs{x}, t)$. Here, $T$ denotes the periodicity, through which the natural frequency $\omega_n$ can be calculated as $\omega_n = 2\pi/T$. The periodic nature of the flow allows for phase $\theta$ to satisfy
\begin{equation}
\dot{\theta} = \omega_n \, , \quad \theta \in [-\pi, \pi] \, . \label{eqn:Phase1}
\end{equation}
Note that $\bs{q_0}\big(\theta(t)\big)$ describes the full flow field at time $t$ corresponding to phase $\theta(t)$. In the vicinity of the limit cycle, and in the basin of the attractor, the phase dynamics can be captured via the phase function $\mathit{\Theta}(\bs{q})$ so that $\theta = \mathit{\Theta}(\bs{q})$. Consequently, we find that  
\begin{equation}  
\dot{\theta} = \mathit{\dot{\Theta}}(\bs{q}) = \bnabla_{\bs{q}} \mathit{\Theta}(\bs{q}) \cdot \dot{\bs{q}} = \bnabla_{\bs{q}} \mathit{\Theta}(\bs{q}) \cdot \bs{F}(\bs{q}) = \omega_n \, , \label{eqn:Phase2}
\end{equation}
which provides a one-dimensional, linear framework for describing the originally nonlinear and high-dimensional dynamics of (\ref{eqn:Dyn1}) with respect to the limit cycle $\bs{q_0}$.      

The flow dynamics of (\ref{eqn:Dyn1}) can then be weakly perturbed in the following fashion
\begin{equation}
\dot{\bs{q}} = \bs{F}(\bs{q})  + \varepsilon \bs{f}(t) \, , \label{eqn:Dyn2}
\end{equation}
where $\varepsilon \ll 1$ and $\norm{\bs{f}} = 1$. This equation can be expressed in terms of the phase variable and function to arrive at 
\begin{equation}  
\dot{\theta} = \mathit{\dot{\Theta}}(\bs{q}) = \bnabla_{\bs{q}} \mathit{\Theta}(\bs{q}) \cdot \dot{\bs{q}} = \bnabla_{\bs{q}} \mathit{\Theta}(\bs{q}) \cdot [\bs{F}(\bs{q}) + \varepsilon \bs{f}(t) ] \, . \label{eqn:Phase_forced}
\end{equation}
Equations (\ref{eqn:Phase_forced}) and (\ref{eqn:Phase2}) can then be combined to obtain
\begin{equation}  
\dot{\theta} = \omega_n + \varepsilon \bnabla_{\bs{q}} \mathit{\Theta}(\bs{q})\big|_{\bs{q} = \bs{q_0}}\cdot\bs{f}(t) \, . \label{eqn:LRF}
\end{equation}
Here, the higher-order terms have been neglected, assuming that the perturbation is sufficiently small. Hereafter, we denote $\bnabla_{\bs{q}} \mathit{\Theta}(\bs{q}) |_{\bs{q} = \bs{q_0}}$ by $\bs{Z}(\theta)$, and refer to it as the phase-sensitvity function.

We find the phase-sensitivity function via a direct method by applying weak impulse perturbations in the form $I \delta(t - t_0) \hat{\bs{e}}_{\bs{j}}$ at different phase values over a period by changing $t_0$, where $I$, $\delta(t-t_0)$ and $\hat{\bs{e}}_{\bs{j}}$ represent the perturbation amplitude, the Dirac delta function centered at $t_0$ and the unit vector specifying the impulse direction, respectively. In this research, $\delta(t-t_0)$ is modelled as a narrow Gaussian function given by 
\begin{equation}  
\delta(t-t_0) = \frac{1}{\sqrt{2\pi} \sigma} \exp\bigg[ - \frac{1}{2}\Big(\frac{t - t_0}{\sigma}\Big)^2 \bigg] \, , \label{eqn:Dirac_delta}
\end{equation}
where $\sigma$ is taken as $10 \Delta t$, with $\Delta t$ indicating the DNS time step (see \S \ref{section:DNS}). 

The introduction of impulse perturbation results in an asymptotic phase advancement or delay. This asymptotic phase shift, known as the phase-response function $g(\theta; I\hat{\bs{e}}_{\bs{j}})$, can be measured experimentally or numerically for a certain phase value and location at which the perturbation is added. Given that $I \ll 1$, the phase-sensitivity function can be evaluated as       
\begin{equation}  
Z_j(\theta) = \lim_{I \rightarrow 0} \frac{g(\theta; I\hat{\bs{e}}_{\bs{j}})}{I} \approx \frac{g(\theta; I\hat{\bs{e}}_{\bs{j}})}{I} \, . \label{eqn:PSF}
\end{equation}    
The phase-sensitivity functions can be computed for various phase values and locations, to obtain the $2\pi$-periodic phase-sensitivity functions of the entire flow field. One can alternatively integrate the adjoint equations of the dynamical system backward in time to obtain $\bs{Z}$ \citep{Ermentrout2010, Nakao2016}. This adjoint-based approach relies on the calculation of Jacobi matrix of $ \bs{F}(\bs{q})$ at each time, which requires the explicit knowledge of governing equations, and as a result, cannot be implemented experimentally.

Once $\bs{Z}$ is available, the linear model of (\ref{eqn:LRF}) can be used for designing optimal open- or closed-loop controls \citep{Nair2020}. It can also be utilized to study the synchronization of the periodic flow to an external actuation such as $\varepsilon \bs{f}(t)$ with the frequency $\Omega$, assuming that $\abs{\Omega - \omega_n} \ll 1$. We define the relative phase between the actuation and the flow oscillation as $\phi(t) = \omega_n - \Omega t$. This definition can be combined with equation (\ref{eqn:LRF}) to calculate the temporal rate of change of $\phi$ 
\begin{equation}  
\dot{\phi} = \varepsilon[\Delta + \bs{Z}(\phi + \Omega t) \cdot \bs{f}(t)] \, , \label{eqn:Relative_phase1}
\end{equation}
where $\Delta = (\omega_n - \Omega)/\varepsilon = \mathcal{O}(1)$. 

The synchronization or phase-locking between the flow oscillation and harmonic forcing occurs when $\dot{\phi} \rightarrow 0$. As a consequence, and after many cycles, the flow adjusts its frequency, so that it oscillates with the new frequency $\Omega$. Since $\Omega t$ changes much more rapidly than the relative phase ($\abs{\Omega - \omega_n} \ll 1$), the right-hand side of equation (\ref{eqn:Relative_phase1}) can be estimated by taking its average over one period \citep{Kuramoto1984, Ermentrout2010}. Hence, equation (\ref{eqn:Relative_phase1}) can be simplified to   
\begin{equation}  
\dot{\phi} = \varepsilon[\Delta + \mathit{\Gamma}(\phi)]  \, , \label{eqn:Relative_phase2}
\end{equation}
where the $2\pi$-periodic phase-coupling function $\mathit{\Gamma}(\phi)$ is formulated as 
\begin{equation}  
\mathit{\Gamma}_m(\phi) = \frac{\Omega}{2\pi} \int_{0}^{2\pi/\Omega} \bs{Z} (\phi + \Omega t'/m ) \cdot \bs{f}(t') \mathrm{d}t' = \frac{1}{2\pi} \int_{0}^{2\pi} \bs{Z}(\phi + \varphi/m) \cdot \bs{f}(\varphi) \mathrm{d}\varphi \, , \label{eqn:PCF}
\end{equation}
while $\Omega = m\omega_n$ with $m \in \mathbb{N}$. Stable phase-locking between the external forcing and the periodic flow can be achieved if and only if
\begin{equation}  
\varepsilon \mathit{\Gamma}_{\mrm{min}} < \Omega - \omega_n < \varepsilon \mathit{\Gamma}_{\mrm{max}} \, ,
\label{eqn:Sync_condition}
\end{equation}
where $\mathit{\Gamma}_{\mrm{max}}$ and $\mathit{\Gamma}_{\mrm{min}}$ are respectively the maximum and minimum values of the phase-coupling function at a certain point. The region of synchronization in the frequency-amplitude plane, also called Arnold tongue \citep{Arnold1997}, is thus bounded by two lines with the slopes $-1/\Gamma_{\mrm{min}}$ and $1/\Gamma_{\mrm{max}}$, both passing through the point $(\Omega/\omega_n, \varepsilon) = (1, 0)$ (e.g., see figure \ref{fig:Arnold_tongue}). Evidently, as the difference between $\mathit{\Gamma}_{\mrm{max}}$ and $\mathit{\Gamma}_{\mrm{min}}$ becomes larger, the area of the synchronization region grows as well. We therefore define the parameter \textit{synchronizability} $S \equiv \mathit{\Gamma}_{\mrm{max}} - \mathit{\Gamma}_{\mrm{min}}$ as a measure for the convenince of phase-locking. While the above derivations assume $\varepsilon \ll 1$, the synchronization criterion (\ref{eqn:Sync_condition}) often holds for moderate values of $\varepsilon$.

\section{Numerical simulations \label{section:DNS}}

\begin{figure}
 \centerline{\includegraphics[width=1.\textwidth]{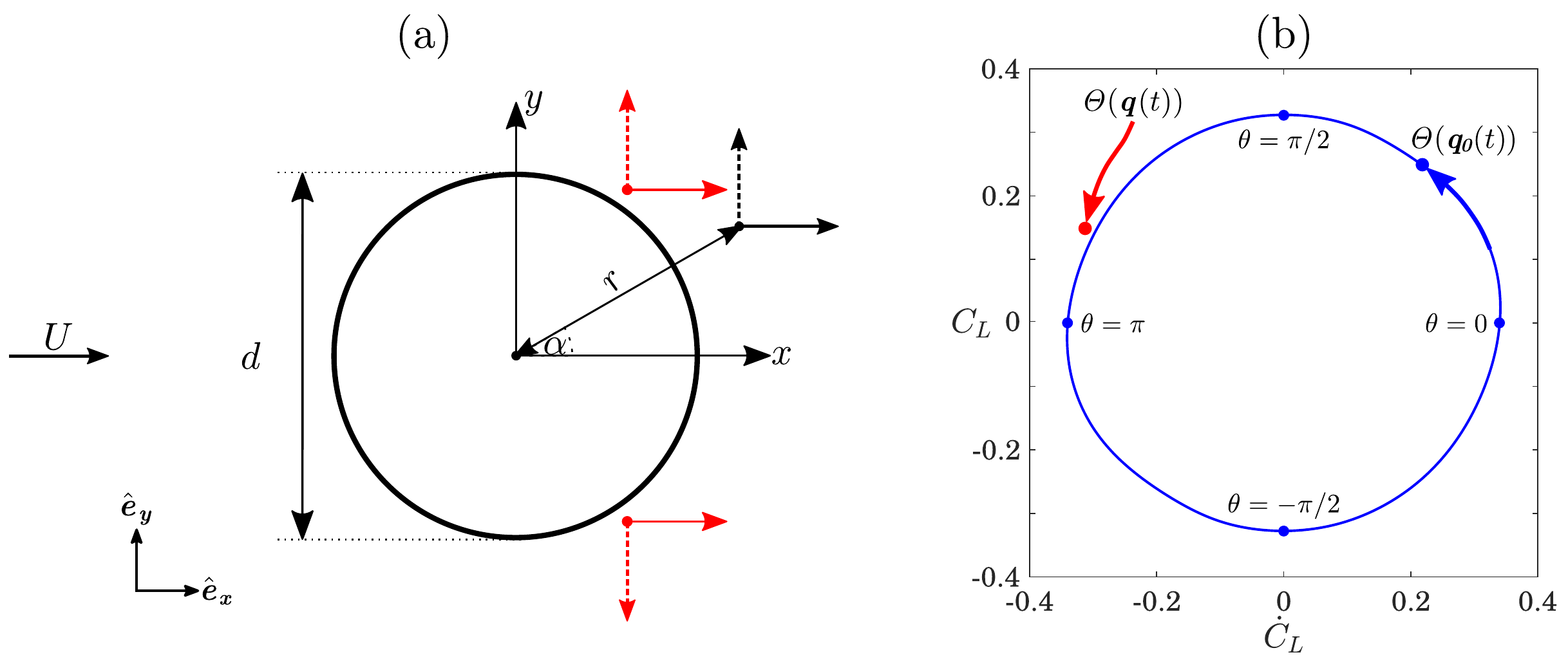}}
\caption{(a) Setup of impulse perturbations and harmonic actuations used for calculating the phase properties of the fluid flow. $\alpha$ and $r$ denote the angular position of the perturbation and its distance from the center of the cylinder, respectively. Horizontal and vertical, black arrows indicate the position and orientation of impulse perturbations used for developing the linear, phase-based model, while the red arrows show example arrangements of top and bottom actuators used to study the synchronization proeprties of the wake flow. (b) Limit-cycle attractor of the flow depicted in the $\dot{C}_L-C_L$ plane, along with the definitions of the phase variable $\theta$ and the phase function $\mathit{\Theta}$.}
\label{fig:Setup}
\end{figure}

Let us consider the application of the phase-reduction analysis to characterize the synchronization properties of the circular cylinder wake. The governing equations for this 2D incompressible flow are
\begin{eqnarray}
\frac{\partial \bs{u}}{\partial t} + \bs{u} \cdot \bnabla{\bs{u}} = \frac{1}{Re} \nabla^2 \bs{u} - \nabla p + \varepsilon \bs{f} \,  \quad \textrm{and} \quad \bnabla \cdot \bs{u} = 0 \, ,
\label{eqn:DNS}
\end{eqnarray}
where $\bs{u} = u \hat{\bs{e}}_{\bs{x}} + v \hat{\bs{e}}_{\bs{y}}$, $p$ and $\bs{f}$ represent dimensionless velocity, pressure and external forcing, respectively. Equations in (\ref{eqn:DNS}) are non-dimensionalized by taking the cylinder diameter $d$ and the freestream velocity $U$ as the characteristic length and velocity (figure \ref{fig:Setup}(a)). Here, we study the flow at a Reynolds number of $Re \equiv Ud/\nu = 100$, where $\nu$ denotes the kinematic viscosity. For the cylinder flow under consideration, the state of the flow is determined by measuring the lift coefficient $C_L$ and its temporal derivative $\dot{C}_L$ (figure \ref{fig:Setup}(b)), whose asymptotic changes in response to the impulse perturbations give the phase response, and subsequently, the phase-sensitivity function of the wake flow.   


A filter in the form of the three-point discrete delta function of \cite{Roma1999} is multiplied by the forcing $\bs{f}$ of equation (\ref{eqn:DNS}) to ensure its spatial compactness. This is enforced in the DNS solver by the following integral over the entire domain
\begin{equation}
     \bs{f}(\bs{x}, t) = \sum_{j = 1}^{n_f} \int_{\bs{x^{\prime}}} \delta_x(\bs{x^{\prime}} - \bs{x}_j)\hat{\bs{e}}_{\bs{j}} T(t) \ \mrm{d} \bs{x^{\prime}}  \, ,  \label{eqn:DNS_force}
\end{equation}
where $n_f$, $\delta_x$ and $\bs{x}_j$ respectively show the number of forcings, the discrete delta function and the location at which the forcings are applied. The temporal part of the forcing $T(t)$ represents the impulse perturbations $I \delta (t - t_0)$ of \S \ref{section:Theory} or the periodic forcings of the next section. Such impulsive, point forces are placed at different angular positions $\alpha$ and distances $r$ from the center of the cylinder, to map out the phase-sensitivity function for the entire flow field (figure \ref{fig:Setup}(a)). The perturbation amplitudes are selected such that $I \in [0.01, 0.02]$, making them adequately small to satisfy the linearity assumption.

The DNS of governing equations (\ref{eqn:DNS}) is carried out using the immersed boundary projection method of \cite{Taira2007}. The far-field boundary conditions are handled via the multi-domain technique of \cite{Colonius2008}, while each domain is discretized uniformly. The computational domain is chosen as $x/d \in [-39.5, 40.5]$ and $y/d \in [-40, 40]$ with the cylinder located at the origin. The first domain, which is the nearest to the cylinder, lies within the area $[-2, 3]\times[-2.5, 2.5]$ and has the grid spacing $\Delta x/d = \Delta y/d = 0.025$. The time step $\Delta t$ is set to satisfy $\Delta t < 0.5\Delta x/U$. When the impulse perturbations are located too close to the boundaries of the first domain or fall outside of it, the domain is sufficiently enlarged. The natural frequency, the time series of lift and drag coefficients, and the velocity and vorticity fields rendered by the present computational approach are in agreement with those of earlier studies such as \cite{Munday2013} and \cite{Taira2018}, validating the overall setup.

\section{Results \label{section:Results}}

In this section, we discuss the results regarding the maps of phase-sensitivity functions and synchronizability for the entire flow field. Such findings reveal valuable information about designing optimal control strategies for the frequency or phase modification of the wake dynamics. The accuracy of the model predictions for the synchronizability are validated against DNS results. We furthermore study the connections and differences between the present method and the DMD-based analysis of the flow, in order to illustrate the distinctive features of the phase-reduction analysis. 

\begin{figure}
  \centerline{\includegraphics[width=0.99\textwidth]{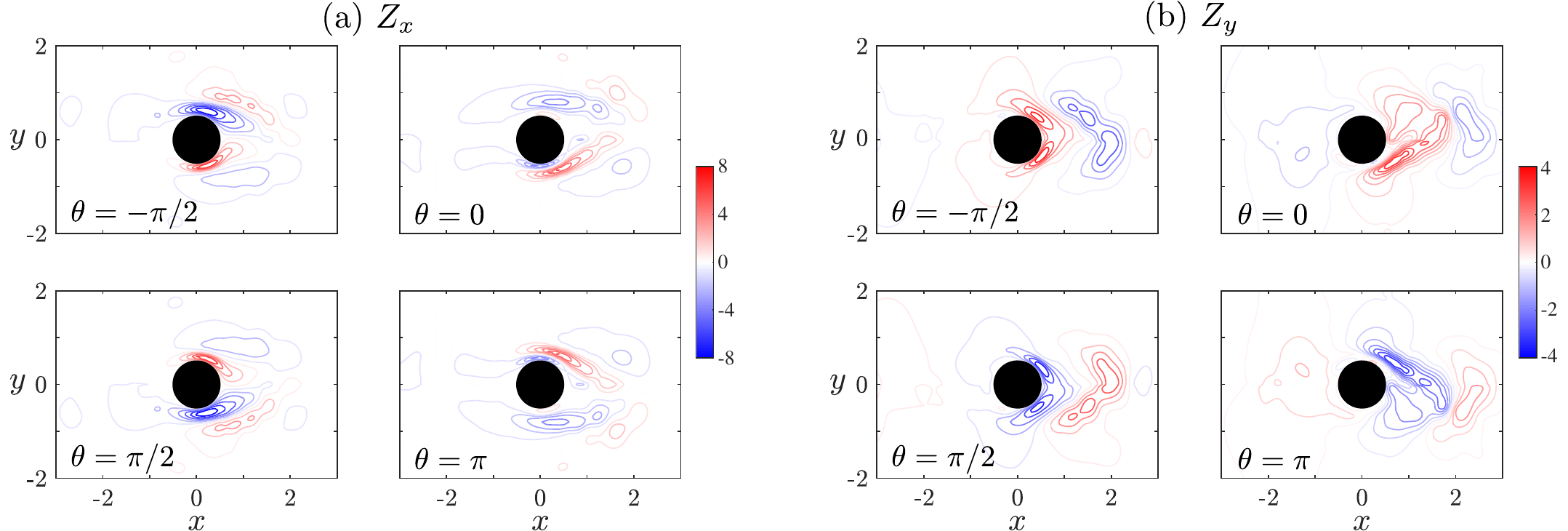}}
  \caption{Phase-sensitivity function at different phases $\theta$, given by impulse perturbations in (a) $\hat{\bs{e}}_{\bs{x}}$ and (b) $\hat{\bs{e}}_{\bs{y}}$ directions.}
\label{fig:Z}
\end{figure}

The spatial maps of phase-sensitivity functions demonstrated in figure \ref{fig:Z} exhibit a symmetry condition between the upper and lower halves of the flow so that for horizontal impulses $Z_x(\theta; x/d, \alpha) = Z_x(\theta + \pi; x/d, -\alpha)$, and for the vertical ones $Z_y(\theta; x/d, \alpha) = -Z_y(\theta + \pi; x/d, -\alpha)$. In accordance with the findings of \cite{Luchini2009} and \cite{Iima2019}, we observe that the sensitivity to the forcing inputs is typically the largest downstream of the cylinder and in its vicinity. In addition, a horizontal perturbation in positive direction and placed upstream of the cylinder, either does not change the phase $\theta$ ($Z_x = 0$) or causes phase advancement ($Z_x < 0$), as also reported by \cite{Iima2019}. The maps of figure \ref{fig:Z} indicate that an energy-efficient control strategy for shifting the phase of the flow requires placing the control inputs around the separation point ($\pi/3 \lesssim \abs{\alpha} \lesssim \pi/2$ and $0.6 \lesssim x/d \lesssim 0.8$), when they are oriented horizontally, and much closer to the wake ($\pi/12 \lesssim \abs{\alpha} \lesssim \pi/4$ and $0.6 \lesssim x/d \lesssim 0.8$), when they are vertical. The stated regions correspond to the peaks of $Z_x$ and $Z_y$.

\begin{figure}
  \centerline{\includegraphics[width=1.\textwidth]{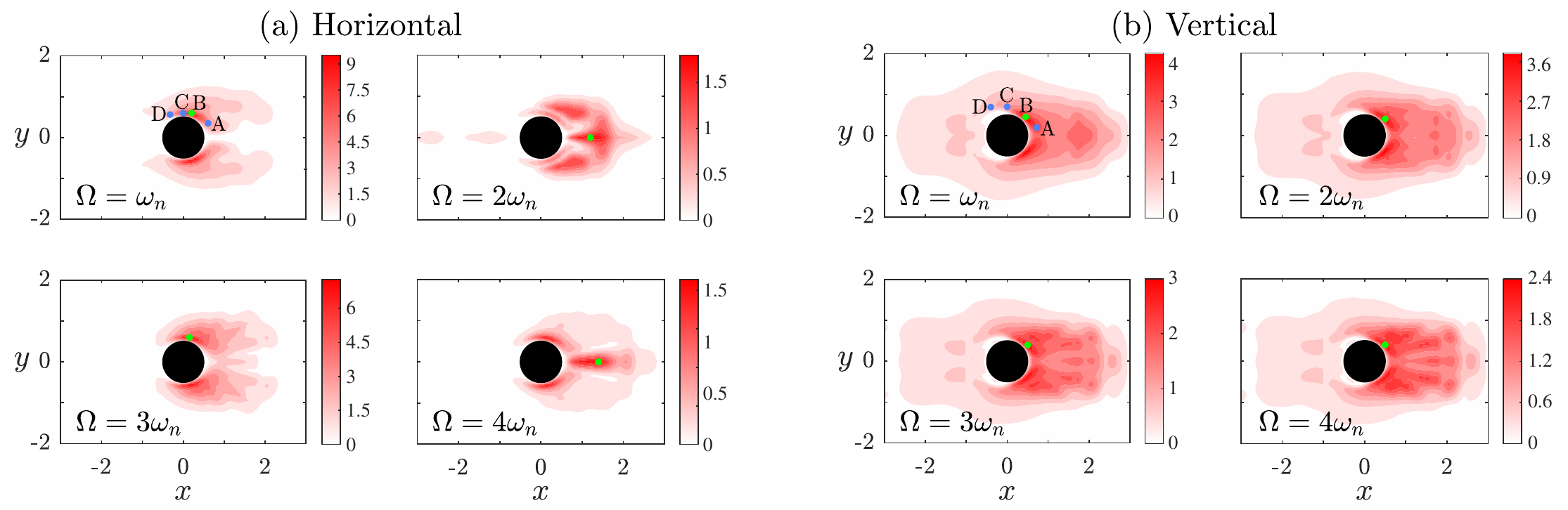}}
  \caption{(a) Synchronizability fields at different harmonics for horizontal periodic actuators. The green point in each panel indicates the position at which $S$ for a certain value of $\Omega/\omega_n$ is maximized. (b) Same as (a), except harmonic actuators are directed vertically.}
\label{fig:Sync}
\end{figure}

\begin{figure}
  \centerline{\includegraphics[width=1.0\textwidth]{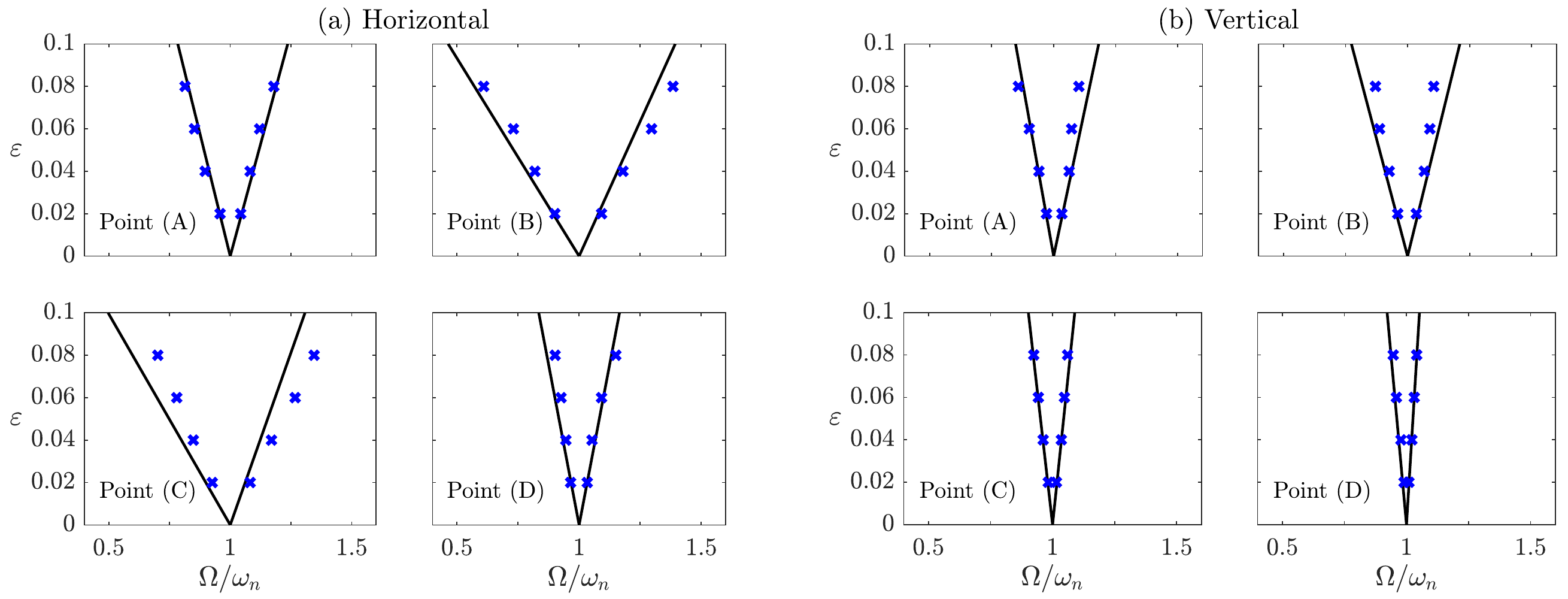}}
  \caption{Synchronization of vortex shedding to harmonic actuations for (a) horizontal and (b) vertical forcing inputs, located at points A to D of figure \ref{fig:Sync}. The phase-based analysis suggests that only cases falling inside the areas bounded by the solid black lines meet the theoretical criteria for phase-locking. The numerically found boundaries of synchronization are also marked by the blue crosses.}
\label{fig:Arnold_tongue}
\end{figure}

In this study, we consider the synchronization of the wake to two sinusoidal forcing inputs in the form $f = 0.5\varepsilon[1 + \sin(\Omega t)]$, placed at the top and bottom of the cylinder, as shown in figure \ref{fig:Setup}. Note also that there is a phase difference $\pi$ between the two actuators. The symmetric distributions of synchronizability $S$ with respect to the wake centerline are depicted at different harmonics in figure \ref{fig:Sync}, when $\bs{f}$ is directed horizontally and vertically. As can be seen in this figure, for the odd-harmonic forcings, i.e. when $\Omega  = (2k-1)\omega_n$ ($k \in \mathbb{N}$), horizontal actuations give much larger values for $S$ in comparison with vertical actuations, suggesting that for these forcings, phase-locking to horizontal actuations can take place with much lower forcing amplitudes. For the even-harmonic forcings, however, the opposite holds, meaning that synchronization to vertical actuations at these harmonics requires lower values for $\varepsilon$. Furthermore, the regions with large synchronizability in the first harmonic mainly overlap with those of the associated phase-sensitivity map. In fact, the location of maximum synchronizability for horizontal (vertical) actuations at $\Omega = \omega_n$ is found to be at $x/d = 0.6$ and $\alpha = 5\pi/12$ ($x/d = 0.6$ and $\alpha = \pi/4$), marked by the green point of the corresponding map. For vertical periodic forcings, the general shape of synchronizability maps does not change substantially for different harmonics, and the maximum value of $S$ appears at the same location for all. In contrast, when the forcings are horizontal, the pattern of $S$ can significantly differ among different harmonics. Specifically, when $\Omega = 2k\omega_n$, $S$ in the centerline of the wake increases from approximately zero at the first harmonic to some fairly large values in $\mathcal{O} (1)$. This can be attributed to the $\pi$-periodicity of $Z_x$ along the centerline of the wake, leading to the orthogonality of phase-sensitivity function to $\bs{f}$, and nearly zero values for both $\mathit{\Gamma}_{\mrm{max}}$ and $\mathit{\Gamma}_{\mrm{min}}$ when $\Omega = \omega_n$. At $\Omega = 2\omega_n$ or other higher harmonics, this orthogonality no longer holds, as can be understood from the formulation for the phase-coupling function (\ref{eqn:PCF}), resulting in relatively large values for $S$ at $\alpha = 0$. In fact, the maximum value of $S$ occurs at $\alpha = 0$ and $x/d = 1.2$ ($x/d = 1.4$) for $\Omega = 2\omega_n$ ($\Omega = 4\omega_n$). The synchronization of actuations placed at the colored points shown in the synchronizability maps of the first harmonic are compared against DNS results in figure \ref{fig:Arnold_tongue}.

\begin{figure}
  \centerline{\includegraphics[width=1.0\textwidth]{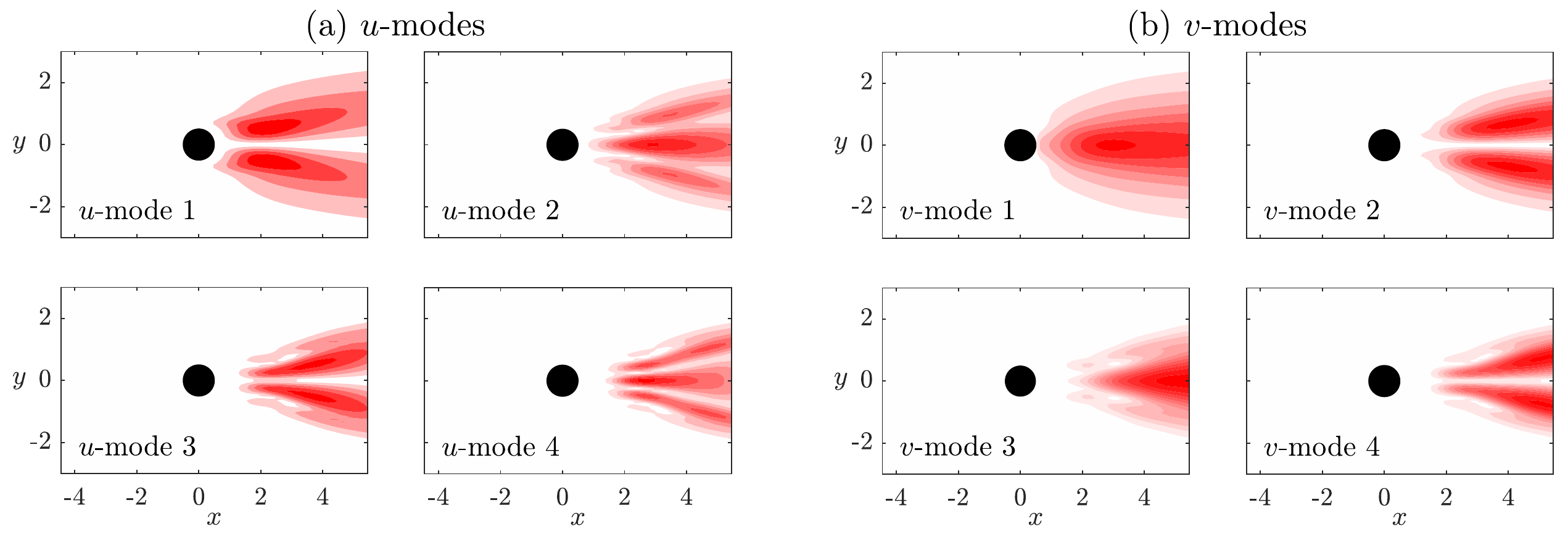}}
  \caption{Magnitude of DMD modes for (a) horizontal, and (b) vertical velocity fields, with the frequency of mode $m$ being $m\omega_n$.}
\label{fig:DMD}
\end{figure}

The comparison between the findings of the present phase-based approach, particularly for synchronizability maps, and the DMD analysis of the wake flow, underlines the characteristic differences between the two methods. Most distinctively, the phase-based model finds the optimal location for the synchronization fairly close to the cylinder, while the peaks of DMD modes fall at least two diameters away from the cylinder, as can be seen for the velocity modes of figure \ref{fig:DMD}. We should emphasize that the phase-reduction and DMD analyses are employed for different objectives. The former identifies the optimal location for an oscillatory control input capable of modifying the phase or oscillation frequency of the flow, whereas the latter aims to extract the most dynamically relevant structures of the flow field. The comparison also exhibits some commonalities between the two. Most noticeably, it shows that the patterns identified by DMD modes are typically strong along the same directions in which $S$ is large. For instance, the structures in DMD modes of the horizontal velocity field are vanishing at $\alpha = 0$ for the first and third modes, but are the strongest for the second and fourth modes. In addition, for the odd modes of vertical velocity, it is observed that the dominant structures emerge closer to the wake center in comparison with those of $u$-modes. These findings are in line with the results of the present analysis for synchronizability fields.

The present phase-based model reveals the necessary conditions for the synchronization of the vortex shedding to external actuations. The theoretically calculated Arnold tongues corresponding to the points A to D of figure \ref{fig:Sync} (solid lines in figure \ref{fig:Arnold_tongue}) are compared against the numerically obtained boundaries of synchronization in the frequency-amplitude plane for the same cases (blue crosses in the same figure). The numerical boundaries for a prescribed value of $\varepsilon$ are found by locating the largest value of $\abs{\Omega - \omega_n}$ for which the difference between the oscillation frequency of the flow and $\Omega$ is below $1\%$. Except for the case with horizontal actuation at $x/d = 0.6$ and $\alpha = \pi/2$ (point C in figure \ref{fig:Sync}a), in which model predictions slightly deviate from DNS results, all other cases show excellent agreement between the two. The linearity of the phase-based model requires a careful parametric study for the perturbation amplitude $I$, to ensure that $I$ is adequately small so that its further decrease does not change the phase-sensitivity function noticeably. This step has significant impact on the overall accuracy of the model predictions. Furthermore, as the amplitude of the excitation grows, the state of the flow moves farther from its stable limit cycle, and as a consequence, nonlinear effects become gradually more pronounced, however, the results of figure \ref{fig:Arnold_tongue} evidence that the model remains accurate for a broad range of forcing amplitudes. Consistent with the observations of \cite{Rigas2017} and \cite{Herrmann2020}, we also find that subharmonic synchronization (i.e., when the vortex shedding oscillates with the frequency $\Omega/m$ instead of $\Omega$) can commonly arise at higher harmonics. However, since the smaller values of $S$ at these harmonics (especially at the even harmonics of horizontal synchronizability field) make them difficult to achieve, phase-locking at higher harmonics are not shown here. 

\begin{figure}
  \centerline{\includegraphics[width=1.0\textwidth]{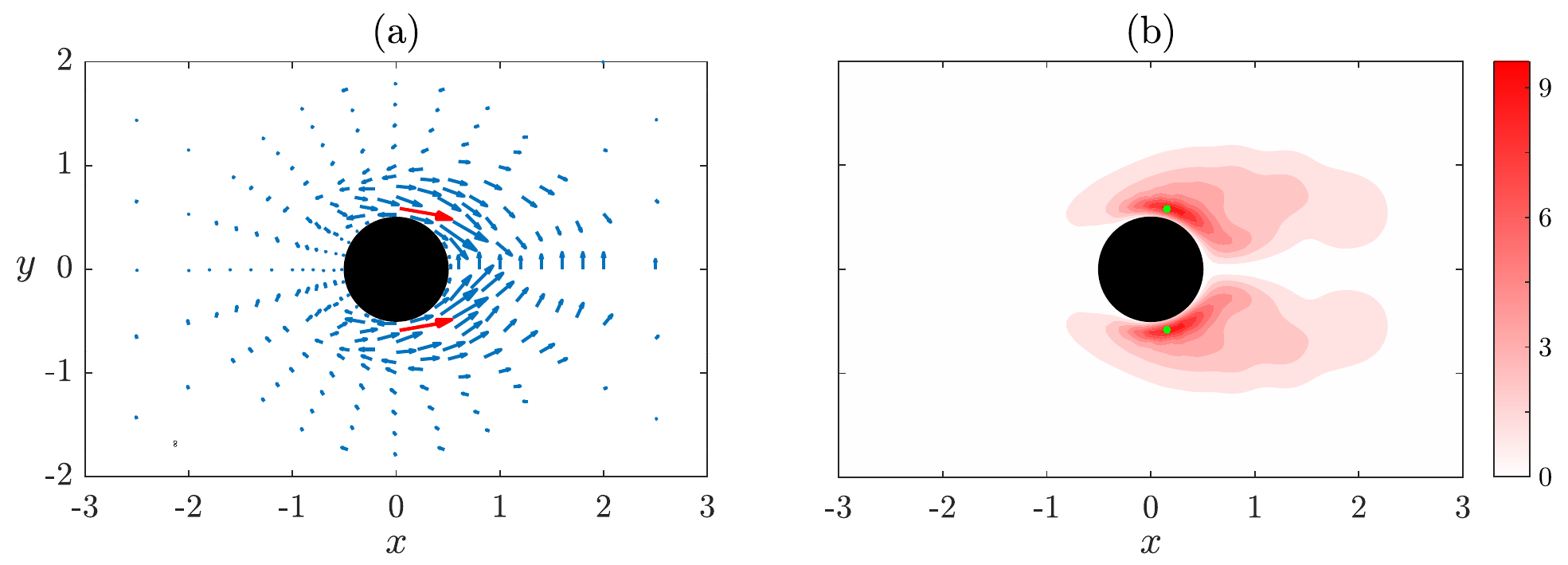}}
  \caption{(a) The vector field represents the optimal orinetation of forcing inputs at different locations in the flow field, in order to acquire the maximum synchronizability. The length of each vector is proportional to the synchronizability achieved by applying the perturbation in the same direction and at the same location as the vector. The red vectors correspond to the actuations leading to the maximum value of $S$. (b) Synchronizability field at $\Omega = \omega_n$, when the periodic actuations are aligned with the top red vector of panel (a).}
\label{fig:Orientation}
\end{figure}

Once the maps of $Z_x$ and $Z_y$ are available, the phase-sensitivity function for any arbitrariliy-oriented impulse perturbation at any point can be calculated as $Z(\beta) = Z_x \cos(\beta) + Z_y \sin (\beta)$, while $\beta$ is the angle between the forcing direction and $\hat{\bs{e}}_{\bs{x}}$. Subsequently, other phase properties such as $\Gamma$ and $S$ can also be computed for all possible locations and directions of forcing inputs, following the formulations of \S \ref{section:Theory}. The optimal orientation of actuations leading to the maximum synchronizability for different points in the flow field are shown in figure \ref{fig:Orientation}. The length of each vector is proportional to the magnitude of $S$, when a harmonic forcing in the same direction and at the same location is introduced. The optimal orientation for the periodic actuations in the near wake ($x/d \lesssim 1.5$) is found to be nearly tangential to the cylinder. This is of practical significance, since tangential forcings can result in a more streamlined and stabilized wake, causing substantial drag reduction \citep{Munday2013}. It is also seen that actuations tangential to the cylinder and placed at $x/d = 0.6$ and $\alpha = \pm 5\pi/12$ (red arrows of figure \ref{fig:Orientation}(a)) yield the largest value possible for $S$. Therefore, a harmonic actuation at the same location as the red vectors and parallel to them results in phase-locking with the smallest amplitude $\varepsilon$, in comparison with all other actuations at any other locations and with any other orientations. The synchronizability field corresponding to periodic forcings parallel to the top red arrow of figure \ref{fig:Orientation}(a), and with $\Omega = \omega_n$, is remarkably similar to the synchronizability map of horizontal forcings at the same harmonic (compare figures \ref{fig:Orientation}(b) and \ref{fig:Sync}(a)), except the values of $S$ in figure \ref{fig:Orientation}(b) are around $5\%$ larger.   

The capabilities of the present phase-based analysis in modifying the oscillation frequency of the flow can give insights into the physics of vortex shedding process as well as the interaction mechanicsms of vortical structures in the wake flow such as vortex splitting and merging. Furthermore, this method proposes a novel approach for controlling the unsteady dynamics of periodic flows through changing their oscillation frequency or phase by finding the optimal strategy for the placement of forcing inputs. Such strategy can in turn enable substantial drag reduction and/or lift enhancement. The current framework of the phase-reduction theory holds promise for its application to a variety of engineering problems such as biological swimmers and flyers, vortex-induced vibration and fluid-structure interactions.      

\section{Concluding Remarks \label{section:Conclusion}}

We performed a phase-based analysis on the 2D periodic flow past a cylinder at $Re = 100$. Weak impulse perturbations were introduced at different locations in the flow field and at various times over a period, in order to uncover the important phase properties of the periodic wake. The present direct method does not need a priori knowledge of the underlying dynamics, is computationally tractable, and can be straightforwardly implemented in experimental setups. The maps of phase-sensitivity function and synchronizability identified by the phase-based model of (\ref{eqn:LRF}) reveal optimal strategies for altering the wake dynamics via modification of the phase or frequency of the flow. The present method accurately predicts the necessary conditions for the synchronization of the wake flow to external harmonic actuations, as confirmed by comparing the model predictions against DNS results. Furthermore, the comparison between the phase- and DMD-based analysis of the flow shows that, despite some interesting similarities, the optimal positioning of forcing inputs required for changing the frequency or the phase of the flow is better determined using the present phase-based approach. This should not come across as surprisng, as they are developed for vastly different objectives. In subsequent studies, the authors aim to investigate the commonalities and discrepancies between the two methods in detail, consider more complex physics such as the interactions between the periodic wakes of multiple bodies, and extend the current approach to quasiperiodic flows, e.g. via cluster-based techniques.      

\section*{Acknowledgment \label{section:Acknowledgment}}

We gratefully acknowledge the support from the US Air Force Office of Scientific Research (Grant: FA9550-16-1-0650, Program Managers: Drs. Gregg Abate and Douglas Smith) and the Army Research Office (Grant: W911NF-19-1-0032, Program Manager: Dr. Matthew J. Munson).


\bibliographystyle{taira}
\bibliography{Main}

\begin{thebibliography}{23}
\expandafter\ifx\csname natexlab\endcsname\relax\def\natexlab#1{#1}\fi
\def\au#1{#1} \def\ed#1{#1} \def\yr#1{#1}\def\at#1{#1}\def\jt#1{\textit{#1}}
  \def\bt#1{#1}\def\bvol#1{\textbf{#1}} \def\vol#1{#1} \def\pg#1{#1}
  \def\publ#1{#1}\def\arxiv#1{#1}\def\org#1{#1}\def\st#1{\textit{#1}}

\bibitem[Arnold(1997)]{Arnold1997}
{\sc \au{Arnold, V.~I.}} \yr{1997} {\em Mathematical Methods of Classical
  Mechanics\/}.  \publ{Springer, New York}.

\bibitem[Bewley(2001)]{Bewley2001}
{\sc \au{Bewley, T.~R.}} \yr{2001}  \at{Flow control: {N}ew challenges for a
  new renaissance}.  \jt{Progress in Aerospace Sciences}  \bvol{37}~(1),
  \pg{21--58}.

\bibitem[Colonius \& Taira(2008)]{Colonius2008}
{\sc \au{Colonius, T.} \& \au{Taira, K.}} \yr{2008}  \at{A fast immersed
  boundary method using a nullspace approach and multi-domain far-field
  boundary conditions}.  \jt{Comput. Methods Appl. Mech. Engrg.}  \bvol{197},
  \pg{2131--2146}.

\bibitem[Ermentrout \& Terman(2010)]{Ermentrout2010}
{\sc \au{Ermentrout, G.~B.} \& \au{Terman, D.~H.}} \yr{2010} {\em Mathematical
  Foundations of Neurosience\/}.  \publ{Springer, New York}.

\bibitem[Herbert {\em et~al.\/}(1987)Herbert, Bertolotti \&
  Santos]{Herbert1987}
{\sc \au{Herbert, T.}, \au{Bertolotti, F.~P.} \& \au{Santos, G.~R.}} \yr{1987}
  Floquet analysis of secondary instability in shear flows.  \bt{In {\em
  Stability of time dependent and spatially varying flows\/}},  \pg{pp.
  43--57}.

\bibitem[Herrmann {\em et~al.\/}(2020)Herrmann, Oswald, Semaan \&
  Brunton]{Herrmann2020}
{\sc \au{Herrmann, B.}, \au{Oswald, P.}, \au{Semaan, R.} \& \au{Brunton,
  S.~L.}} \yr{2020}  \at{Modeling synchronization in forced turbulent
  oscillator flows}.  \jt{arXiv:2001.12000} .

\bibitem[Holmes {\em et~al.\/}(1996)Holmes, Lumley \& Berkooz]{Holmes1996}
{\sc \au{Holmes, P.}, \au{Lumley, J.~L.} \& \au{Berkooz, G.}} \yr{1996} {\em
  Turbulence, coherent structures, dynamical systems and symmetry\/}.
  \publ{Cambridge University Press, Cambridge}.

\bibitem[Huerre \& Monkewitz(1990)]{Huerre1990}
{\sc \au{Huerre, P.} \& \au{Monkewitz, P.~A.}} \yr{1990}  \at{Local and global
  instabilities in spatially developing flows}.  \jt{Annu. Rev. Fluid Mech.}
  \bvol{22},  \pg{473--537}.

\bibitem[Iima(2019)]{Iima2019}
{\sc \au{Iima, M.}} \yr{2019}  \at{Jacobian-free algorithm to calculate the
  phase sensitivity function in the phase reduction theory and its applications
  to {K}\'{a}rm\'{a}n's vortex street}.  \jt{Phys. Rev. E}  \bvol{99},
  \pg{062203}.

\bibitem[Kawamura \& Nakao(2013)]{Kawamura2013}
{\sc \au{Kawamura, Y.} \& \au{Nakao, H.}} \yr{2013}  \at{Collective phase
  description of oscillatory convection}.  \jt{Chaos}  \bvol{23},  \pg{043129}.

\bibitem[Kawamura \& Nakao(2015)]{Kawamura2015}
{\sc \au{Kawamura, Y.} \& \au{Nakao, H.}} \yr{2015}  \at{Phase description of
  oscillatory convection with a spatially translational mode}.  \jt{Physica D}
  \bvol{295-296},  \pg{11--29}.

\bibitem[Kuramoto(1984)]{Kuramoto1984}
{\sc \au{Kuramoto, Y.}} \yr{1984} {\em Chemical oscillations, waves, and
  turbulence\/}.  \publ{Springer, Berlin}.

\bibitem[Kuramoto \& Nakao(2019)]{Kuramoto2019}
{\sc \au{Kuramoto, Y.} \& \au{Nakao, H.}} \yr{2019}  \at{On the concept of
  dynamical reduction: {T}he case of coupled oscillators}.  \jt{Philosophical
  Transactions of the Royal Society A}  \bvol{377},  \pg{20190041}.

\bibitem[Luchini {\em et~al.\/}(2009)Luchini, Giannetti \&
  Pralits]{Luchini2009}
{\sc \au{Luchini, J.}, \au{Giannetti, F.} \& \au{Pralits, J.~O.}} \yr{2009}
  Structural sensitivity of the finite-amplitude vortex shedding behind a
  circular cylinder.  \bt{In {\em Proc. IUTAM Symp. Unsteady Sep. Flows
  Control\/}},  \pg{pp. 151--160}.

\bibitem[Munday \& Taira(2013)]{Munday2013}
{\sc \au{Munday, P.~M.} \& \au{Taira, K.}} \yr{2013}  \at{On the lock-on of
  vortex shedding to oscillatory actuation around a circular cylinder}.
  \jt{Phys. Fluids}  \bvol{25},  \pg{013601}.

\bibitem[Nair {\em et~al.\/}(2018)Nair, Brunton \& Taira]{Nair2018}
{\sc \au{Nair, A.~G.}, \au{Brunton, S.~L.} \& \au{Taira, K.}} \yr{2018}
  \at{Networked-oscillator-based modeling and control of unsteady wake flows}.
  \jt{Phys. Rev. E}  \bvol{97},  \pg{063107}.

\bibitem[Nair {\em et~al.\/}(2020)Nair, Taira, Brunton \& Brunton]{Nair2020}
{\sc \au{Nair, A.~G.}, \au{Taira, K.}, \au{Brunton, B.~W.} \& \au{Brunton,
  S.~L.}} \yr{2020}  \at{Phase-based control of fluid flows}.
  \jt{arXiv:2004.10561} .

\bibitem[Nakao(2016)]{Nakao2016}
{\sc \au{Nakao, H.}} \yr{2016}  \at{Phase reduction approach to synchronisation
  of nonlinear oscillators}.  \jt{Contemporary Phys.}  \bvol{57}~(2),
  \pg{188--214}.

\bibitem[Pikovsky {\em et~al.\/}(2001)Pikovsky, Rosenblum \&
  Kurths]{Pikovsky2001}
{\sc \au{Pikovsky, A.}, \au{Rosenblum, M.} \& \au{Kurths, J.}} \yr{2001} {\em
  Synchronization: A universal concept in nonlinear sciences\/}.
  \publ{Cambridge University Press, Cambridge}.

\bibitem[Rigas {\em et~al.\/}(2017)Rigas, Morgans \& Morrison]{Rigas2017}
{\sc \au{Rigas, G.}, \au{Morgans, A.~S.} \& \au{Morrison, J.~F.}} \yr{2017}
  \at{Weakly nonlinear modelling of a forced turbulent axisymmetric wake}.
  \jt{J. Fluid Mech.}  \bvol{814},  \pg{570--591}.

\bibitem[Roma {\em et~al.\/}(1999)Roma, Peskin \& Berger]{Roma1999}
{\sc \au{Roma, A.~M.}, \au{Peskin, C.~S.} \& \au{Berger, M.~J.}} \yr{1999}
  \at{An adaptive version of the immersed boundary method}.  \jt{J. Comput.
  Phys.}  \bvol{153},  \pg{509--534}.

\bibitem[Taira \& Colonius(2007)]{Taira2007}
{\sc \au{Taira, K.} \& \au{Colonius, T.}} \yr{2007}  \at{The immersed boundary
  method: {A} projection approach}.  \jt{J. Comput. Phys.}  \bvol{225},
  \pg{2118--2137}.

\bibitem[Taira \& Nakao(2018)]{Taira2018}
{\sc \au{Taira, K.} \& \au{Nakao, H.}} \yr{2018}  \at{Phase-response
  synchronization for analysis of periodic flows}.  \jt{J. Fluid Mech.}
  \bvol{846},  \pg{R2}.

\end{thebibliography}

\end{document}